\documentclass[twocolumn]{autart}

\usepackage{amsmath}
\usepackage{amssymb}
\usepackage[tight,footnotesize]{subfigure}
\usepackage{balance}
\usepackage{paralist}

\usepackage{ifthen}
\newboolean{jour}%
\setboolean{jour}{true}
\newcommand{\jour}[1]{\ifthenelse{\boolean{jour}}{#1}{}}
\newcommand{\conf}[1]{\ifthenelse{\boolean{jour}}{}{#1}}

\newboolean{jouralt}%
\setboolean{jouralt}{true}
\newcommand{\jouralt}[1]{\ifthenelse{\boolean{jouralt}}{#1}{}}
\newcommand{\confalt}[1]{\ifthenelse{\boolean{jouralt}}{}{#1}}

\newboolean{testcut}%
\setboolean{testcut}{true}
\newcommand{\testcut}[1]{\ifthenelse{\boolean{testcut}}{#1}{}}
\newcommand{\notcut}[1]{\ifthenelse{\boolean{testcut}}{}{#1}}

\renewcommand{\phi}{\varphi}
\renewcommand{\epsilon}{\varepsilon}

\renewcommand{\qed}{\hfill $\square$}

\newcommand{\bcd}{\theta}

\newtheorem{theorem}{\bf Theorem}

\newtheorem{ass}[theorem]{\bf Assumption}

\usepackage{epic,eepic,psfrag}
\usepackage{pgf}

\usepackage{verbatim}

\newcommand{\Xcon}{\mathbb{X}}
\newcommand{\Xest}{\mathbb{X}}
\newcommand{\Xfun}{\mathbb{X}}

\newcommand{\Yest}{\mathbb{Y}}
\newcommand{\Yfun}{\mathbb{Y}}

\newcommand{\Uest}{\mathbb{U}}
\newcommand{\Ufun}{\mathbb{U}}

\newcommand{\Vcon}{\mathbb{V}}
\newcommand{\Vest}{\mathbb{V}}
\newcommand{\Vfun}{\mathbb{V}}

\newcommand{\Wcon}{\mathbb{W}}
\newcommand{\West}{\mathbb{W}}
\newcommand{\Wfun}{\mathbb{W}}

\newcommand{\met}{|}

\newcommand{\Nall}{2}
\newcommand{\Ndelta}{\Nall}
\newcommand{\Ngamma}{\Nall}

\newcommand{\plus}{\oplus}
\newcommand{\Plus}{\bigoplus}

\newcommand{\plusTria}{\plus}

\newcommand{\TriaConst}[1]{2}

\begin{document}

\begin{frontmatter}

\title{Nonlinear Full Information and Moving Horizon Estimation: Robust Global Asymptotic Stability\thanksref{label1}}
\thanks[label1]{This work was supported by the German Research Foundation under Grant MU3929-2/1, project number: 426459964.
}

\author[address_irt,address_bosch]{Sven Kn\"ufer}
and
\ead{knuefer@irt.uni-hannover.de}
\author[address_irt]{Matthias A. M\"uller}
\ead{mueller@irt.uni-hannover.de}

\address[address_irt]{Institute of Automatic Control, Leibniz University Hannover, 30167 Hannover, Germany.}
\address[address_bosch]{Robert Bosch GmbH, Driver Assistance, 70469 Stuttgart, Germany.}

\begin{abstract}
In this paper, we propose time-discounted schemes for full information estimation (FIE) and moving horizon estimation (MHE) that are robustly globally asymptotically stable (RGAS).
We consider general nonlinear system dynamics with nonlinear process and output disturbances that are a priori unknown.
For FIE being RGAS, our only assumptions are that the system is time-discounted incrementally input-output-to-state-stable (i-IOSS) and that the time-discounted FIE cost function is compatible with the i-IOSS estimate.
Since for i-IOSS systems such a compatible cost function can always be designed, we show that i-IOSS is sufficient for the existence of RGAS observers.
Based on the stability result for FIE, we provide sufficient conditions such that the induced MHE scheme is RGAS as well for sufficiently large horizons.
For both schemes, we can guarantee convergence of the estimation error in case the disturbances converge to zero without incorporating a priori knowledge.
Finally, we present explicit converge rates and show how to verify that the MHE results approach the FIE results for increasing horizons.
\end{abstract}

\begin{keyword}
Moving horizon estimation \sep Full information estimation \sep Robust stability \sep Nonlinear systems \sep Detectability
\end{keyword}

\end{frontmatter}

\section{Introduction} \label{sec_intro}

Many practical control applications require an estimate of the internal system state, for instance to use state-feedback control algorithms or to monitor a safe and efficient operation.
In such practical environments, state estimators need to handle nonlinear system dynamics and to guarantee robustness against unknown process and measurement disturbances.
To this end, optimization-based state estimators such as full information estimators (FIE) or moving horizon estimators (MHE) gained increasing attention in the recent years, see~\cite{Allan_Diss2020,Gharbi_Ebenbauer_2019,Gharbi_Ebenbauer_2020,Rawlings_Ji_JPC12,Rawlings_Mayne_Diehl_MPC17,Wynn_et_al_TAC14}.
In FIE, an optimization problem is used at each time instant to estimate a trajectory that reproduces all previous output measurements with a minimal deviation from the nominal system dynamics measured by a stage cost function that penalizes the size of the corresponding process and output disturbances.
The end point of this trajectory serves as state estimate at the current time instant.
Since FIE takes all output measurements into account, it becomes computationally intractable with increasing time.
In MHE, this issue is resolved by taking into account only a fixed number of the most recent output measurements and by penalizing the distance to a previous state estimate in the MHE cost function.
Due to this setup, FIE and MHE can naturally address nonlinear system dynamics and allow to include knowledge about constraints on the system states or disturbances. %
A particular strength of MHE and FIE is that guarantees for robustness can be shown even in this nonlinear case.

Early results on FIE and MHE handle observable systems, see~\cite{Alessandri_et_al_Aut08,Alessandri_et_al_CDC10,Michalska_Mayne_1992,Muske_et_al_ACC_1993}, or undisturbed systems or disturbances that are a priori known to converge to zero over time, see~\cite{Rao_et_al_TAC03,Rawlings_Ji_JPC12,Rawlings_Mayne_09}.
In~\cite{Hu_et_al_CDC15,Ji_et_al_MHE}, FIE is shown to be robustly asymptotically stable in presence of bounded disturbances by adding a max-term in the cost function.
However, convergence of the estimation error in case of converging disturbance could not be shown without incorporating according a priori knowledge in the optimization problem.
For MHE, the same approach allowed to show robust asymptotic stability and convergence in~\cite{Hu_arXiv_2017,Muller_ACC16}, where, however, the resulting disturbance gains are not shown to improve with increasing horizons but remain constant.
In~\cite{Muller_Aut_2017}, MHE is proven to be robustly asymptotically stable and convergent even without the additional max-term in the cost function, i.e., using a classical (weighted) least-squares cost function, but the resulting disturbance gains even increase, i.e., get worse, with a larger horizon length.
All results~\cite{Hu_arXiv_2017,Hu_et_al_CDC15,Ji_et_al_MHE,Muller_ACC16,Muller_Aut_2017,Rao_et_al_TAC03,Rawlings_Ji_JPC12,Rawlings_Mayne_09} have in common that incremental input-output-to-state stability (i-IOSS) conditions are used to ensure nonlinear detectability.
Under an exponentially time-discounted i-IOSS condition, FIE is shown to be robustly globally exponentially stable (RGES) in~\cite{knuefer2018} by using exponentially time-discounted stage costs.
Additionally assuming a global Lipschitz condition for the system dynamics,\footnote{Note that systems which are exponentially i-IOSS are guaranteed to have globally Lipschitz system dynamics if the output function ${h}$ is globally Lipschitz.} RGES is proven also for MHE and convergence rates and disturbance gains are presented that converge towards the ones for FIE in case the horizon length is increased, see~\cite{knuefer2018}.
Introducing an incremental stabilizability condition with respect to the process disturbances, the authors of~\cite{Allan_Rawlings_ACC2019} define a Lyapunov-like so-called Q-function for the estimation error, which allows to analyze ${\mathcal{KL}}$-stability for FIE.
In~\cite{Allan_Diss2020}, this approach is extended to the analysis of RGAS for FIE and it is shown that exponentially stabilizable systems satisfying an exponentially decaying i-IOSS condition admit FIE and MHE schemes which are RGES.

In the same line as the stability results for optimization-based state estimation have improved in recent years, also the underlying detectability conditions have evolved.
While different notions are used, all publications~\cite{Hu_arXiv_2017,Hu_et_al_CDC15,Ji_et_al_MHE,knuefer2018,knuefer2020,Muller_ACC16,Muller_Aut_2017,Rao_et_al_TAC03,Rawlings_Ji_JPC12,Rawlings_Mayne_09} cited above rely on or handle i-IOSS.
Initially, i-IOSS has been introduced and shown to be necessary for the existence of RGAS observers in~\cite{Sontag_OSS}.
In order to verify that a system satisfies the i-IOSS condition, the authors of~\cite{Allan_Rawlings_ACC2019} present a Lyapunov-like condition for the system dynamics with additive output disturbances.
For the time-discounted version that~\cite{Allan_Diss2020,Allan_Rawlings_ACC2019} are based on, equivalence to non time-discounted i-IOSS is shown in~\cite{AllanRawlingsTeel_SIAM_2021} and a converse theorem is formulated stating that a system is i-IOSS if and only if an i-IOSS Lyapunov function exists.
In~\cite{knuefer2020}, the notion of i-IOSS is extended to general nonlinear system dynamics and disturbances, and Lyapunov-like conditions are presented for this setting that allow to verify the sum-based or the max-based i-IOSS condition.
In summary, i-IOSS represents a notion of detectability that is a) necessary for the existence of stable observers, b) covering general system setups, and c) verifiable via Lyapunov-like conditions. %

The main contributions of this work are as follows.
While the previous results on FIE and MHE handle additive output disturbances, we consider general system dynamics with a priori unknown nonlinear process and output disturbances in the present work.
For detectability, the time-discounted i-IOSS condition discussed in~\cite{knuefer2020} is supposed to hold.
In addition, a compatibility condition for the i-IOSS estimate and the cost function is used that can always be guaranteed for i-IOSS systems by designing the cost function accordingly.
For this setup, we show RGAS for FIE for bounded disturbances without introducing an additional stabilizability condition, without implicitly assuming an exponential i-IOSS condition, and without incorporating a priori knowledge about the disturbances.
To the authors' best knowledge, the FIE scheme presented in this work is the first RGAS observer that effectively relies on an i-IOSS condition only, c.f.~\cite[Section~6.2.1]{Allan_Diss2020}.
Hence, we show that i-IOSS is not only necessary but also sufficient for the existence of RGAS observers.
Moreover, explicit convergence rates are presented without requiring incremental stabilizability as, e.g., in~\cite{Allan_Diss2020,Allan_Rawlings_ACC2019}.
Furthermore, this work presents sufficient conditions such that FIE induces an RGES or RGAS MHE scheme if the horizon length is chosen sufficiently large.
To this end, the stability result for FIE is required to provide a not necessarily linear contraction map that is iterated in the MHE case.
A similar approach based on a \emph{linear} contraction map has been used in~\cite{Allan_Diss2020,Hu_arXiv_2017,Muller_Aut_2017}.
However, our arguments to verify that FIE induces such a contraction map implicitly require an \emph{eventually\footnote{Naming according to~\cite{Allan_Rawlings_MHE2019}.}} exponential i-IOSS condition as of now, see Remark~\ref{rem:MHEcontractionImplicatons} in Section~\ref{sec:MHE} for a detailed discussion.
As for FIE, explicit convergence rates and disturbance gains are presented for MHE.
Finally, we show that these rates approach the ones for FIE for increasing horizons.

The present work is structured as follows.
In Section~\ref{sec_preliminaries_setup}, the notation and setup is presented.
Moreover, i-IOSS is introduced as detectability condition and the cost function for the FIE and MHE optimization problem is defined.
Sections~\ref{sec:FIE} and~\ref{sec:MHE} show and discuss our stability results for FIE and MHE, respectively.
Finally, we conclude our work with Section~\ref{sec:conclusions}.

\section{Preliminaries and Setup}
\label{sec_preliminaries_setup}

Let
${\Xfun}$,
${\Ufun}$,
${\Yfun}$,
${0 \in \Wfun}$,
and
${0 \in \Vfun}$
be metric spaces with corresponding metrics ${\met{} \cdot, \cdot \met{}}$ and abbreviate ${\met{} \cdot, 0\met{}}$ by ${\met{} \cdot \met{}}$.\footnote{Using metric spaces allows to emphasize that none of the following steps requires vector space structure. The element ${0}$ of ${\Wfun}$ and ${\Vfun}$ is simply referred to as a nominal representative and is a rather arbitrary choice in this setup.}
In the following, we consider nonlinear discrete-time system dynamics of the form
\begin{align}
	x(t+1) &= f( x(t), u(t), w(t) ),
  \label{eq:sys}
	\\
	y(t) &= h( x(t), u(t), v(t) ),
  \label{eq:out}
\end{align}
where ${t \in \mathbb{N} \ (\ni 0)}$ and where ${f: \Xfun \times \Ufun \times \Wfun \rightarrow \Xfun}$, ${h: \Xfun \times \Ufun \times \Vfun \rightarrow \Yfun}$ are some nonlinear functions constituting the system dynamics and the output model, respectively.
In \eqref{eq:sys}-\eqref{eq:out}, ${u : \mathbb{N} \rightarrow \Ufun}$ denotes the known control input, ${w : \mathbb{N} \rightarrow \Wcon}$ represents an a priori unknown process disturbance, and ${v : \mathbb{N} \rightarrow \Vcon}$ defines an a priori unknown measurement noise.
An initial condition ${x_0 \in \Xcon}$, an input ${u}$, and a process disturbance ${w}$ lead to a state trajectory ${x: \mathbb{N} \rightarrow \Xcon}$ under \eqref{eq:sys}.
Finally, the measurement noise ${v}$ generates an output trajectory ${y: \mathbb{N} \rightarrow \Yfun}$ according to~\eqref{eq:out}.
Such a tuple ${ \{ x, u, w, v, y \} }$ satisfying \eqref{eq:sys}-\eqref{eq:out} for all ${t \in \mathbb{N}}$ is called a solution of system~\eqref{eq:sys}-\eqref{eq:out} in the following.
For ${K \in \mathbb{N} \cup \{\infty\}}$, let ${\Sigma^{K} \subset \Xest^{K} \times \Uest^{K} \times \West^{K} \times \Vest^{K} \times \Yest^{K} }$ denote the set of all such solutions with length ${K}$.

For our detectability condition, for the cost function, and for the stability results, we will make use of the well-known notion of comparison functions according to the following definition.

\begin{defn}[Comparison Functions]
\label{defn:comparisonfun}
A function ${\kappa: [0, \infty) \rightarrow [0, \infty)}$ is called ${\mathcal{K}}$-\emph{function}, i.e., ${\kappa \in \mathcal{K}}$, if ${\kappa}$ is continuous, strictly increasing, and ${\kappa(0)=0}$.
If ${\kappa \in \mathcal{K}}$ is unbounded, it is called ${\mathcal{K}_{\infty}}$-\emph{function}, i.e., ${\kappa \in \mathcal{K}_{\infty}}$. 
A function ${\kappa: \mathbb{N} \rightarrow [0, \infty)}$ is called ${\mathcal{L}}$-\emph{function}, i.e., ${\kappa \in \mathcal{L}}$, if ${\kappa}$ is nonincreasing and ${\lim_{t\rightarrow\infty} \kappa(t)=0}$.
A function ${\kappa: [0, \infty) \times \mathbb{N} \rightarrow [0, \infty)}$ is called ${\mathcal{KL}}$-\emph{function}, i.e., ${\kappa \in \mathcal{KL}}$, if ${\kappa(\cdot, t)} \in \mathcal{K}$ for each fixed ${t \in \mathbb{N}}$, and $\kappa(r,\cdot) \in \mathcal{L}$ for each fixed ${r \in [0, \infty)}$.
A ${\mathcal{KL}}$-function ${\kappa}$ is called \emph{summable} if there exists a bounding ${\mathcal{K}}$-function $\sigma$ such that ${\sum _{\tau = 0} ^{\infty} \kappa( r, \tau ) \leq \sigma(r)}$ holds for all ${r \in \mathbb{R}}$.
\end{defn}

As discussed in the introduction, i-IOSS is an established detectability condition especially in the context of optimization-based state estimation.
However, several notions of i-IOSS have been utilized in this context such as discounted and non time-discounted versions and max-based and sum-based formulations.
While the present work focuses on a time-discounted i-IOSS condition, it also addresses both the max-based and the sum-based notion alike.
To this end, the placeholder ${\plus}$ is used.

\begin{defn}[Placeholder ${\plus}$]
\label{defn:placeholder}
Throughout this work, ${\plus}$ is a placeholder a) for either the summation according to
\vspace{-0.15cm}
\begin{align}
  a \plus b &:= ( a + b )
  \quad
  \text{and}
  \quad
  \Plus _{i = K_1} ^{K_2} a_{i} := \sum_{i = K_1} ^{K_2} a_{i}
\end{align}
or b) for the maximum operation according to
\begin{align}
  a \plus b &:= \max \{ a, b \}
  \quad
  \text{and}
  \quad
  \Plus _{i = K_1} ^{K_2} a_{i} := \max_{K_1 \leq i \leq K_2} a_{i}
\end{align}
for all ${a, b, a_{i} \in \mathbb{R}}$ and ${K_1, K_2 \in \mathbb{N}}$.
Furthermore, let the ${\plus}$ operator always be applied after multiplication, i.e.,
\begin{align}
  a_1 a_2 \plus a_3 a_4 &= (a_1 a_2) \plus (a_3 a_4).
\end{align}
\end{defn}

\begin{rem}
\label{rem:placeholder}
Note that Definition~\ref{defn:placeholder} allows for two interpretations of the placeholder ${\plus}$.
However, the represented operation needs to be chosen for the entire argumentation of the present work and not locally on a term-by-term basis.
Precisely, the reader can exchange the ${\plus}$-symbol \emph{everywhere} in this paper by \emph{either} ${\max}$ \emph{or} ${+}$.
To compare the implications of this global choice, the terms \emph{max-based} and \emph{sum-based} formulations are used.
In all arguments below, only such modifications are applied that hold for both operations alike.
(In the limited cases where a concrete operation is needed or intended we explicitly keep the ${\max}$ or ${+}$ operation.)
Although the two operations represented by ${\plus}$ share certain properties such as associativity and commutativity, maximization and summation are in general not exchangeable of course.
In this context, an important difference is that maximization is distributive with respect to ${\mathcal{K}}$-functions, while summation is not, i.e.,
${\kappa(\max \{ a, b \}) = \max \{ \kappa(a), \kappa(b) \}}$
and
${\kappa( a + b ) \neq \kappa(a) + \kappa(b)}$.
\end{rem}

The notion introduced in Definition~\ref{defn:comparisonfun} and \ref{defn:placeholder} allows to formulate time-discounted i-IOSS, which serves as detectability condition in the following.

\begin{defn}[Time-Discounted i-IOSS, see~\cite{knuefer2020}]
\label{defn:iIOSSsummable}
System \eqref{eq:sys}-\eqref{eq:out} is time-discounted incrementally input-output-to-state stable (i-IOSS) if there exist ${\alpha \in \mathcal{K}_{\infty}}$ and ${\beta, \gamma, \delta, \epsilon, \varphi \in \mathcal{KL}}$ such that, for any two solutions ${ \{ x, u, w, v, y \} }$ and ${ \{ \chi, \upsilon, \omega, \nu, \zeta \} }$ of~\eqref{eq:sys}-\eqref{eq:out}, the difference between the two trajectories remains bounded according to
\begin{align}
\label{eq:defn:iIOSSsummable}
	\alpha(|x(t), \chi(t)|)
  \hspace{0.5cm}&\hspace{-0.5cm}
  \leq \beta( |x_{0}, \chi_{0}|, t )
	\\
	&
  \notag
  \plus \Plus _{\tau = 1} ^{t} ( \gamma( |w(t - \tau), \omega(t - \tau)|, \tau )
	\\
	& \qquad
  \notag
  \plus \delta( |v(t - \tau), \nu(t - \tau)|, \tau )
	\\
	& \qquad
  \notag
  \plus \epsilon( |u(t - \tau), \upsilon(t - \tau)|, \tau )
	\\
	& \qquad
  \notag
  \plus \varphi( |y(t - \tau), \zeta(t - \tau)|, \tau ) )
\end{align}
for all ${t \in \mathbb{N}} $ %
and with summable ${\gamma, \delta, \epsilon, \varphi}$ in the sum-based case.
\end{defn}

The above definition using the ${\plus}$ placeholder allows for a max-based and a sum-based formulation.
The max-based formulation is rather established in the literature.
Though, slight differences are to be noted.
As discussed in the introduction, most previous works~\cite{Hu_arXiv_2017,Hu_et_al_CDC15,Ji_et_al_MHE,Muller_ACC16,Muller_Aut_2017,Rao_et_al_TAC03,Rawlings_Ji_JPC12,Rawlings_Mayne_09} on optimization-based state estimation refer to similar max-based notions of i-IOSS as detectability condition.
In the initially proposed form of i-IOSS in~\cite{Sontag_OSS}, the disturbance terms occur in a non time-discounted max-based fashion.
However, the time-discounted formulation even appears to be the one that naturally results from a related Lyapunov condition, see~\cite{Allan_Rawlings_ACC2019,knuefer2020}.
This Lyapunov condition is not only sufficient for i-IOSS, but also necessary for a large class of systems (${\Xfun \subseteq \mathbb{R}^{n}}$, ${h = h(x)}$) as recently shown in~\cite{AllanRawlingsTeel_SIAM_2021}.
For this class, it is shown in~\cite{AllanRawlingsTeel_SIAM_2021} that the max-based time-discounted and the non time-discounted version are equivalent, which implies that the max-based and the sum-based formulation of i-IOSS according to Definition~\ref{eq:defn:iIOSSsummable} are qualitatively equivalent.\footnote{
On the one hand, note that any max-based i-IOSS system is also sum-based i-IOSS as, by a suitable choice of ${\alpha}$, all ${\mathcal{KL}}$-functions in Definition~\ref{defn:iIOSSsummable} can be upper-bounded by ${\mathcal{KL}}$-functions that decrease exponentially in their second arguments and are hence summable, see~\cite[Proposition~7]{Sontag_1998}.
On the other hand, note that any sum-based i-IOSS system with summable ${\mathcal{KL}}$-functions implies a non time-discounted max-based estimate to hold, which is equivalent to a max-based time-discounted i-IOSS result for the aforementioned class of systems.}
Finally, the above i-IOSS condition according to Definition~\ref{defn:iIOSSsummable} is known to be necessary for the existence of RGAS observers, see~\cite{AllanRawlingsTeel_SIAM_2021,knuefer2020}.

For FIE and MHE, candidate trajectories are evaluated in terms of their deviation from the nominal system dynamics.
To this end the following cost function is used.

\begin{defn}[Time-Discounted Cost Function]
\label{defn:costsummable}
Let ${K, t \in \mathbb{N}}$ with ${t \geq K \geq 1}$, some prior ${\bar{x}_{t-K} \in \Xest}$, and ${\hat{\beta}, \hat{\gamma}, \hat{\delta} \in \mathcal{KL}}$ be given and let ${\hat{\gamma}}$ and ${\hat{\delta}}$ be summable in the sum-based case.
For ${\chi_{t-K} \in \hat{\mathbb{X}}}$, ${\omega \in \West^{K}}$ and, ${\nu \in \Vest^{K}}$ with
${\omega = \{\omega(t-K), ..., \omega(t-1)\}}$
and
\linebreak ${\nu = \{\nu(t-K), ..., \nu(t-1)\}}$
define
\begin{align}
\label{eq:cost_K}
	J_{K}(\chi_{t-K}, \omega, \nu)
  &:= \hat{\beta}(|\chi_{t-K}, \bar{x}_{t-K}|, K )
	\\
  \notag
	& \hspace{-0.5cm}
  \plus \Plus _{\tau = 1} ^{K} ( \hat{\gamma}(| \omega(t - \tau) |, \tau )
	\plus \hat{\delta}(| \nu(t - \tau) |, \tau ) ).
\end{align}
\end{defn}

As for the i-IOSS condition, the place holder ${\plus}$ allows to cover a max-based and a sum-based formulation alike.
Note that in the sum-based case, this formulation covers (time-discounted) least squares type cost functions as a special case.
Using a max-based non time-discounted i-IOSS condition, previous works, e.g., \cite{Hu_arXiv_2017,Hu_et_al_CDC15,Ji_et_al_MHE,Muller_ACC16}, simultaneously incorporated max- and sum-terms in their non time-discounted cost functions in order to address bounded disturbances.
Harmonizing the detectability condition and the cost function by using max-terms in both cases allowed for the progress in terms of bounded disturbances compared to earlier results.
However, the mismatch given by the sum-terms in the cost function and the lack of discounting do not allow to show a) convergence for convergent disturbances for FIE (without a priori knowledge about the disturbances), see~\cite{Hu_et_al_CDC15,Ji_et_al_MHE}, and b) that the convergence gains improve for increasing horizons for MHE, see~\cite{Hu_arXiv_2017,Muller_ACC16}.
In~\cite{knuefer2018}, under additional assumptions, these problems are overcome by using the same form for both the i-IOSS condition and the cost function.
Independent of whether a sum-based or a max-based cost function is desired, it appears favorable to use an i-IOSS condition of the same form to arrive at an according stability result.
Note that this connection does not arise when an additional incremental stabilizability condition is used as in~\cite{Allan_Diss2020,Allan_Rawlings_ACC2019}.

All following results suppose the considered system to be i-IOSS and the cost function to be compatible with the i-IOSS condition in terms of the following assumption, i.e., in space and in time.

\begin{ass}
\label{ass:compatible}
System~\eqref{eq:sys}-\eqref{eq:out} is time-discounted i-IOSS according to Definition~\ref{defn:iIOSSsummable}.
Furthermore, there exists a constant ${B \in (0, \infty)}$ such that the ${\mathcal{KL}}$-functions of Definition~\ref{defn:iIOSSsummable} and Definition~\ref{defn:costsummable} satisfy
\begin{align}
  \label{eq:ass:compatible:beta}
  \beta(\TriaConst{s} r, s) &\leq B \hat{\beta}(r, s),
  \\
  \label{eq:ass:compatible:gamma}
  \gamma(\Ngamma r, s) &\leq B \hat{\gamma}(r, s),
  \\
  \label{eq:ass:compatible:delta}
  \delta(\Ndelta r, s) &\leq B \hat{\delta}(r, s)
\end{align}
for all ${s \in [0, \infty)}$ and all ${r \in [0, \infty)}$.
\end{ass}

\begin{rem}
\label{rem:MainAssumption}
Assumption~\ref{ass:compatible} consists of two parts.
Firstly, it requires the system to be time-discounted i-IOSS, i.e., gives a detectability assumption.
Secondly, it poses a compatibility condition~\eqref{eq:ass:compatible:beta}-\eqref{eq:ass:compatible:delta} between the comparison functions of the detectability condition~\eqref{eq:defn:iIOSSsummable} and of the cost function~\eqref{eq:cost_K} in both coordinates.
Note that choosing ${\hat{\beta}(r, s) := \beta(\TriaConst{s} r, s)}$, ${\hat{\gamma}(r, s) := \gamma(\Ngamma r, s)}$, and ${\hat{\delta}(r, s) := \delta(\Ndelta r, s)}$ always guarantees for equality to hold with ${B = 1}$ in~\eqref{eq:ass:compatible:beta}-\eqref{eq:ass:compatible:delta} of Assumption~\ref{ass:compatible}.
These suggested definitions for ${\hat{\beta}}$, ${\hat{\gamma}}$, and ${\hat{\delta}}$ appear to be the \emph{optimal} choices considering the specific i-IOSS condition.
However, all larger, in the sum-based case summable, ${\mathcal{KL}}$-functions satisfying~\eqref{eq:ass:compatible:beta}-\eqref{eq:ass:compatible:delta} are alternative candidates.
Hence, the second compatibility part of Assumption~\ref{ass:compatible} can always be satisfied by a proper design of the cost function.
To this end, the ${\mathcal{KL}}$-functions of the i-IOSS condition do not necessarily need to be known exactly. %
\end{rem}

Definitions~\ref{defn:iIOSSsummable} and~\ref{defn:costsummable} ensure that the detectability condition and the cost function are of the same structure.
Due to the additional compatibility condition~\eqref{eq:ass:compatible:beta}-\eqref{eq:ass:compatible:delta} of Assumption~\ref{ass:compatible}, we have the following preliminary result that allows to bound the difference between two trajectories in terms of the initial difference and of the disturbance corresponding to the trajectory with the larger cost.

\begin{prop}
\label{prop:decrease}
Suppose Assumption~\ref{ass:compatible} applies and let ${A \in [1, \infty)}$ be arbitrary.
For any ${t, K \in \mathbb{N}}$, ${t \geq K \geq 1}$, let ${ \{ x, u, w, v, y \}, \{ \hat{x}, u, \hat{w}, \hat{v}, y \} \in \Sigma^{K} }$ be two solutions on the time interval ${\{t-K, \dots, t-1\}}$ with the same input and output.
If these two solutions satisfy
\begin{align}
	\label{eq:prop:suboptimal}
	J_{K}(\hat{x}(t-K), \hat{w}, \hat{v})
  \leq
	A J_{K}(x(t-K), w, v),
\end{align}
then the estimate
\begin{align}
	\label{eq:prop:decrease}
	\alpha(|x(t), \hat{x}(t)|)
  &\leq b( | x(t-K), \bar{x}_{t-K} |, K )
	\\
	& \hspace{-0.5cm}
  \notag
  \plus \Plus _{\tau = 1} ^{K} ( c( | w(t - \tau) |, \tau )
	\plus d( | v(t - \tau) |, \tau ) )
\end{align}
holds with ${b, c, d \in \mathcal{KL}}$ according to
\begin{align}
  \label{eq:abbrev:b}
  b(r, s) &:= \beta( \TriaConst{s} r, s ) \plusTria A B \hat{\beta}( r, s )
  \\
  \label{eq:abbrev:c}
  c(r, s) &:= \gamma( \Ngamma r, s ) \plusTria A B \hat{\gamma}( r, s )
  \\
  \label{eq:abbrev:d}
  d(r, s) &:= \delta( \Ndelta r, s ) \plusTria A B \hat{\delta}( r, s )
\end{align}
for all ${s \in [0, \infty)}$ and all ${r \in [0, \infty)}$.%
\end{prop}

\textbf{Proof:}
Since the system is time-discounted i-IOSS according to Definition~\ref{defn:iIOSSsummable} and since the compared trajectories share the same inputs and outputs, we conclude
\begin{align}
	\alpha(|x(t), \hat{x}(t)|)
  \hspace{-1.5cm}&\hspace{1.5cm}
  \leq
  \label{eq:prop:iIOSS}
  \beta( | x(t-K), \hat{x}(t-K) | , K )
	\\
  \notag
	& \plus \Plus _{\tau = 1} ^{K} \gamma( | w(t - \tau), \hat{w}(t - \tau) | , \tau )
	\\
  \notag
	& \plus \Plus _{\tau = 1} ^{K} \delta( | v(t - \tau), \hat{v}(t - \tau) | , \tau )
  \\
  & \hspace{-0.5cm} \leq
  \label{eq:prop:Compatib}
  [ \beta( \TriaConst{K} | x(t-K), \bar{x}_{t-K} |, K )
	\\
  \notag
  & \plusTria B \hat{\beta}( | \hat{x}(t-K), \bar{x}_{t-K} |, K ) ]
	\\
  \notag
	& \plus \Plus _{\tau = 1} ^{K} [
  \gamma( \Ngamma | w(t - \tau) | , \tau )
  \plusTria B \hat{\gamma}( | \hat{w}(t - \tau) |, \tau ) ]
	\\
  \notag
	& \plus \Plus _{\tau = 1} ^{K} [
  \delta( \Ndelta | v(t - \tau) |, \tau )
  \plusTria B \hat{\delta}( | \hat{v}(t - \tau) |, \tau ) ],
\end{align}
where we applied the triangle-inequality
and
the compatibility condition in Assumption~\ref{ass:compatible}.
Due to the definition of the cost-function and to condition~\eqref{eq:prop:suboptimal}, we furthermore obtain
\begin{align}
  \label{eq:prop:Optimality}
  \alpha(|x(t), \hat{x}(t)|)
  \hspace{0.5cm}&\hspace{-0.5cm}\leq
  [ \beta( \TriaConst{K} | x(t-K), \bar{x}_{t-K} |, K )
	\\
  \notag
	& \plus \Plus _{\tau = 1} ^{K} \gamma( \Ngamma | w(t - \tau) |, \tau )
	\\
  \notag
	& \plus \Plus _{\tau = 1} ^{K} \delta( \Ndelta | v(t - \tau) |, \tau ) ]
  \\
  \notag
  & \plusTria A B J_{K}(x(t-K), w, v),
\end{align}
which is equivalent to the stated inequality~\eqref{eq:prop:decrease}.
\qed

\begin{rem}
\label{rem:identicalInputsOutputs}
In the above proof's first step, the i-IOSS estimate~\eqref{eq:defn:iIOSSsummable} is applied and the ${\epsilon}$- and ${\varphi}$-terms are directly eliminated because the considered trajectories have identical inputs and outputs.
As usual, the optimization problems for FIE and MHE will require these identities as additional constraints.
In order to investigate robustness against violations of these constraints, the general i-IOSS formulation of~\eqref{eq:defn:iIOSSsummable} also allows to consider trajectories with deviant inputs, ${u}$ and ${\hat{u}}$, and outputs, ${y}$ and ${\hat{y}}$, in Proposition~\ref{prop:decrease}.
In this case additional terms for the considered constraint violations result in~\eqref{eq:prop:decrease} which can be interpreted and addressed as additional disturbances.
\end{rem}

\begin{rem}
\label{rem:NonDiscounted}
Note that the proof of Proposition~\ref{prop:decrease} does not touch the second argument of the ${\mathcal{KL}}$-functions for the disturbance gains at all.
Hence, the disturbance terms could also be considered in a non time-discounted fashion, e.g., by choosing ${\gamma(r, s) = \gamma(r, 0)}$ and ${\delta}$, ${\hat{\gamma}}$, and ${\hat{\delta}}$ accordingly, in the i-IOSS condition, the cost function, and consequently in~\eqref{eq:prop:decrease}, as long as the compatibility condition in Assumption~\ref{ass:compatible} is satisfied.
Consequently, ${c}$ and ${d}$ of Proposition~\ref{prop:decrease} are no longer converging to zero with respect to the second argument in this case.
The implications for FIE and MHE of this non time-discounted formulation are discussed below in Remarks~\ref{rem:FIENonDiscounted} and~\ref{rem:MHENonDiscounted}, respectively.
\end{rem}

\section{Full Information Estimation}
\label{sec:FIE}

Taking all past measurements into account, the full information estimator defined below makes use of the time-discounted cost function according to Definition~\ref{defn:costsummable}.

\begin{defn}[FIE]
\label{defn:FIE}
Let some initial estimation ${\hat{x}_{0|0} \in \Xest}$ and ${A \in [1, \infty)}$ arbitrary be given.
For ${t \in \mathbb{N}}$, consider the input signals ${u_{[0, t-1]} := \{u(0), \dots, u(t-1) \}}$, the output measurements ${y_{[0, t-1]} := \{y(0), \dots, y(t-1) \}}$, and the prior ${\bar{x}_{0} := \hat{x}_{0|0}}$ to define ${J_{t}}$ according to Definition~\ref{defn:costsummable}.
Determine the estimated state trajectory ${ \hat{x}_{\cdot|t} \in \Xest^{t}}$ and the estimated disturbances ${\hat{w}_{\cdot|t} \in \West^{t}}$, ${\hat{v}_{\cdot|t} \in \Vest^{t}}$ such that
\begin{align}
	J_{t}(\hat{x}_{0|t}, \hat{w}_{\cdot|t}, \hat{v}_{\cdot|t})
  \leq
	&\inf_{ \{ \chi,\omega, \nu \} }
  A J_{t}(\chi(0), \omega, \nu)
	\label{eq:suboptimal_fie}
  \\
  \notag
  &\{ \chi, u_{[0, t-1]}, \omega, \nu, y_{[0, t-1]} \} \in \Sigma^{t}
\end{align}
and ${ \{ \hat{x}_{\cdot|t}, u_{[0, t-1]}, \hat{w}_{\cdot|t}, \hat{v}_{\cdot|t}, y_{[0, t-1]} \} \in \Sigma^{t} }$ are satisfied.
For all ${t \in \mathbb{N}}$, define ${\hat{x}(t) := \hat{x}_{t|t}}$ to be the estimate for the state at time ${t}$.
\end{defn}

At each time instant ${t}$, FIE optimizes over all trajectories that comply with the system dynamics and reproduce the measured outputs.
This optimization allows to relate the costs of the estimated state trajectory ${\hat{x}_{\cdot|t}}$ to the costs of the actual state trajectory ${x}$ according to condition~\eqref{eq:prop:suboptimal} in Proposition~\ref{prop:decrease}.
Hence, the above FIE is RGAS as stated in the following theorem.
Note that choosing ${A = 1}$ requires to find optimal solutions in~\eqref{eq:suboptimal_fie} and factors ${A > 1}$ allow to cover robustness against only suboptimal solutions,\footnote{The proportional suboptimality concept via the factor $A$ allows to define the separate functions ${b}$, ${c}$, and ${d}$ in~\eqref{eq:abbrev:b}-\eqref{eq:abbrev:d}, as distributivity w.r.t. ${\plus}$ is guaranteed.
Alternatively, suboptimality can be represented by a non-linear function as done in~\cite{Hu_arXiv_2017}.
} see~\eqref{eq:abbrev:b}-\eqref{eq:abbrev:d} in Proposition~\ref{prop:decrease}.

\begin{thm}[RGAS of FIE]
\label{thm:FIERGAS}
Suppose Assumption~\ref{ass:compatible} applies.
Then the FIE is RGAS,\footnote{This RGAS formulation is a generalization of the exponential formulation in \cite[Definition~1]{knuefer2018}.
The estimate~\eqref{eq:FIERGAS} implies that converging disturbances result in converging estimates.
It has also been used in the max-based formulation in \cite[Definition~2.2]{Allan_Diss2020}.} i.e., the resulting state estimate satisfies
\begin{align}
	\label{eq:FIERGAS}
	\alpha(|x(t), \hat{x}(t)|)
  &\leq b( | x(0), \hat{x}(0) |, t )
	\\
	& \hspace{-0.5cm}
  \notag
  \plus \Plus _{\tau = 1} ^{t} ( c( | w(t - \tau) |, \tau )
	\plus d( | v(t - \tau) |, \tau ) )
\end{align}
for all ${t \in \mathbb{N}}$, all ${\hat{x}_{0|0} \in \Xest}$, ${x_0 \in \Xcon}$, ${w \in \Wcon^{\infty}}$, and ${v \in \Vcon^{\infty}}$.
\end{thm}

\textbf{Proof:}
As motivated above, (sub-)optimality in \eqref{eq:suboptimal_fie} allows to relate the cost of the estimated state trajectories and the actual one by
\begin{align}
	J_{t}(\hat{x}_{0|t}, \hat{w}_{\cdot|t}, \hat{v}_{\cdot|t})
  \leq
  A J_{t}(x(0), w, v)
\end{align}
for all time instants ${t \in \mathbb{N}}$.
Hence, Proposition~\ref{prop:decrease} with ${K = t}$ directly results in \eqref{eq:FIERGAS} because the prior is chosen according to ${\bar{x}_{0} = \hat{x}_{0|0} = \hat{x}(0)}$.
\qed

Note that for each time instant ${t}$, \eqref{eq:FIERGAS} bounds the estimation error in terms of the initial estimation error and in terms of the actual process and output disturbance.
Moreover, the result implies that the estimation error converges to zero if the disturbances do so and it provides explicit convergence rates.
Hence, Theorem~\ref{thm:FIERGAS} provides an RGAS result in the sense discussed in~\cite{Allan_Rawlings_ACC2019,Allan_Rawlings_MHE2019}.

The essential condition in Theorem~\ref{thm:FIERGAS} is the i-IOSS condition, see Remark~\ref{rem:MainAssumption}.
As discussed in the introduction, this is the first time that a state observer is shown to be RGAS based effectively only on an i-IOSS condition.
Hence, i-IOSS can be considered as necessary and sufficient condition for the existence of RGAS observers.

The placeholder ${\plus}$ allows to present the above RGAS result in the max-based and the sum-based version alike.
For each version however, the i-IOSS condition, the cost function, and the stability result need to be formulated in the same fashion, i.e., all max-based or all sum-based.
A minor difference is given by the fact, that the ${\mathcal{KL}}$-functions for the i-IOSS disturbance gains need to be summable in the sum-based case.
This is a rational addition for the detectability condition as discussed in~\cite{knuefer2020} to ensure finite bounds for bounded disturbances.
Following Remark~\ref{rem:MainAssumption} and Proposition~\ref{prop:decrease},
the cost function can always be designed such that also the relevant ${\mathcal{KL}}$-functions in the stability result are summable.

\begin{rem}
\label{rem:FIENonDiscounted}
As discussed in Remark~\ref{rem:NonDiscounted}, the disturbance terms do not necessarily need to be time-discounted for Proposition~\ref{prop:decrease} and the same applies for Theorem~\ref{thm:FIERGAS}.
For such a non time-discounted version, the functions ${c}$ and ${d}$ are in general not decreasing with respect to time in~\eqref{eq:FIERGAS}.
In this case,~\eqref{eq:FIERGAS} does no longer guarantee that the estimation error converges to zero if the disturbances do so.
This limitation has also been observed in~\cite{Hu_et_al_CDC15,Ji_et_al_MHE} (without incorporating a priori knowledge about the disturbances), where non time-discounted costs including max-terms were used.
Omitting time-discounting of the disturbance terms in the sum-based formulation leads to an unbounded right-hand side in~\eqref{eq:FIERGAS} also for bounded disturbances.
However, this approach might still induce stability for MHE as discussed in Remark~\ref{rem:MHENonDiscounted}.
\end{rem}

\section{Moving Horizon Estimation}
\label{sec:MHE}

Since the FIE optimization problem increases in complexity with increasing time, MHE considers only a fixed number ${K}$ of the most recent output measurements.
In order to make use of the stability result for FIE, the same optimization problem is essentially used in the following definition.

\begin{defn}[MHE]
\label{defn:MHE}
Let some horizon ${K \in \mathbb{N}}$, ${K \geq 1}$, some initial estimation ${\hat{x}_{0|0} \in \Xest}$, and ${A \in [1, \infty)}$ arbitrary be given.
For ${t \in \mathbb{N}}$, ${1 \leq t \leq K}$, apply the FIE scheme according to Definition~\ref{defn:FIE}.
For ${t > K}$, consider the input signals ${u_{[t-K, t-1]} := \{u(t-K), \dots, u(t-1) \}}$, the output measurements ${y_{[t-K, t-1]} := \{y(t-K), \dots,}$ ${y(t-1) \}}$, and the prior ${\bar{x}_{t-K} := \hat{x}_{t-K|t-K}}$\footnote{While we use this so-called filtering prior, different choices are possible, see \cite[Section~4.3.2]{Rawlings_Mayne_09} for a discussion and note that \cite[Assumption~V.1.]{Hu_arXiv_2017} needs to be considered for alternative choices.} to define ${J_{K}}$ according to Definition~\ref{defn:costsummable}.
Determine the estimated state trajectory ${ \hat{x}_{\cdot|t} \in \Xest^{K}}$ and the estimated disturbances ${\hat{w}_{\cdot|t} \in \West^{K}}$, ${\hat{v}_{\cdot|t} \in \Vest^{K}}$ by solving essentially the same optimization problem~\eqref{eq:suboptimal_fie} as for time instant ${K}$, i.e.,  such that
\begin{align}
  \hspace{-0.1cm}
	J_{K}(\hat{x}_{t-K|t}, \hat{w}_{\cdot|t}, \hat{v}_{\cdot|t})
  \leq
	&\inf_{ \{ \chi, \omega, \nu \} }
  A J_{K}(\chi(t-K), \omega, \nu)
  \hspace{-0.08cm}
	\label{eq:suboptimal_mhe}
  \\
  \notag
  &
  \hspace{-1.0cm}
  \{ \chi, u_{[t-K, t-1]}, \omega, \nu, y_{[t-K, t-1]} \} \in \Sigma^{K}
\end{align}
and ${ \{ \hat{x}_{\cdot|t}, u_{[t-K, t-1]}, \hat{w}_{\cdot|t}, \hat{v}_{\cdot|t}, y_{[t-K, t-1]} \} \in \Sigma^{K} }$ are satisfied.
For all ${t \in \mathbb{N}}$, define ${\hat{x}(t) := \hat{x}_{t|t}}$ to be the estimate for the state at time ${t}$.
\end{defn}

All results in the following only depend on the structure of the FIE stability result of Theorem~\ref{thm:FIERGAS}, but in fact there is no direct dependency on the structure of either the i-IOSS condition or the cost function.
Hence, the following results apply to all MHE schemes that are induced by FIE schemes (according to Definition~\ref{defn:MHE}), e.g., induced by FIE as presented in~\cite{Allan_Rawlings_MHE2019}, and that satisfy~\eqref{eq:FIERGAS}.
This general idea of proof, namely to show stability for MHE based on the stability result of the underlying FIE has been presented in \cite{Hu_arXiv_2017} for a cost function containing sum- and max-terms alike and in \cite{Allan_Diss2020} in case FIE is RGES.
Here, we generalize this approach to the asymptotic case for both the sum-based and the max-based formulation.
To this end, the horizon ${K}$ of MHE needs to be chosen large enough such that the stability result~\eqref{eq:FIERGAS} evaluated for ${t=K}$ provides a (not necessarily linear) contraction for the estimation error.
Our first theorem provides a robust global asymptotic stability result based on a contraction for which, in the sum-based case, a relaxed distributivity inequality and a summability condition needs to hold.

\begin{thm}[RGAS of MHE]
\label{thm:MHERGASmax}
Suppose FIE satisfies~\eqref{eq:FIERGAS}.
If for the horizon length ${K}$ there exists ${\kappa_{K} \in \mathcal{K}}$ such that\footnote{\label{fnt:RGASCondMaxBased} Note that \eqref{eq:MHERGASsubdistri}-\eqref{eq:MHERGASsummable} are always satisfied in the max-based case and in case \eqref{eq:MHERGASmaxcond} is satisfied for ${\kappa_{K}}$ linear, see Remark~\ref{rem:MHERGES}.
Additionally, note that \eqref{eq:MHERGASsummable} only serves to ensure that ${\hat{c}_{K}}$ and ${\hat{d}_{K}}$ are summable in the sum-based case, i.e., that \eqref{eq:MHERGASmax} remains bounded for bounded disturbances.}
\begin{align}
  \label{eq:MHERGASmaxcond}
  b( \alpha^{-1}(r) , K ) &\leq \kappa_{K}(r) <  r
  \\
  \label{eq:MHERGASsubdistri}
  \kappa_{K}(r \plus \bar{r}) &\leq \kappa_{K}(r) \plus \kappa_{K}(\bar{r})
  \\
  \label{eq:MHERGASsummable}
  \Plus _{\tau = 0} ^{\infty} \kappa_{K}^\tau (r) &< \infty
\end{align}
holds for all ${r, \bar{r} \in (0, \infty)}$, then MHE is RGAS in the following sense.
The state estimate resulting from MHE satisfies
\begin{align}
	\label{eq:MHERGASmax}
  \hspace{0.5cm}&\hspace{-0.5cm}
	\alpha(|x(t), \hat{x}(t)|)
  \leq
  \hat{b}_{K}( | x(0), \hat{x}(0) |, t )
	\\
	&
  \notag
  \plus \Plus _{\tau = 1} ^{t} (
  \hat{c}_{K}( | w(t - \tau) |, \tau )
  \plus \Plus _{\tau = 1} ^{t}
  \hat{d}_{K}( | v(t - \tau) |, \tau )
  )
\end{align}
for all ${t \in \mathbb{N}}$, all ${\hat{x}_{0|0} \in \Xest}$, ${x_0 \in \Xcon}$, ${w \in \Wcon^{\infty}}$, and ${v \in \Vcon^{\infty}}$ with ${\hat{b}_{K}, \hat{c}_{K}, \hat{d}_{K} \in \mathcal{KL}}$ according to
\begin{align}
  \label{eq:DefHatThetaK}
  \hat{\theta}_{K}(r, t) &:=
  \max \{ \kappa_{K}^{\lfloor t / K \rfloor} ( \theta( r, t \bmod K ) ),
  \\
  \notag
  & \hspace{1.35cm}
  \kappa_{K}^{\lfloor t / K \rfloor + 1} ( \theta( r, 0 ) ) \}
\end{align}
for ${(\theta, \hat{\theta}_{K}) \in \{(b, \hat{b}_{K}), (c, \hat{c}_{K}), (d, \hat{d}_{K})\}}$.
\end{thm}

\textbf{Proof of Theorem~\ref{thm:MHERGASmax}:}
Let ${e(t) := \alpha(|x(t), \hat{x}(t)|)}$ and ${t = k K + l}$ with ${k, l \in \mathbb{N}}$ and minimal ${l}$, i.e., ${l = t \bmod K}$ and ${k = \lfloor t / K \rfloor}$.
Since the MHE optimization problem is derived from FIE, we can directly apply \eqref{eq:FIERGAS} of Theorem~\ref{thm:FIERGAS} and~\eqref{eq:MHERGASmaxcond} to obtain
\begin{align}
  \label{eq:MHEInducStep}
	e(t)
  \hspace{0.5cm}&\hspace{-0.5cm}
  \leq
  \kappa_{K} (e(t-K))
	\\
	& %
  \notag
  \plus
	\Plus _{\tau = 1} ^{K} (
  c( | w(t - \tau) |, \tau )
  \plus
  d( | v(t - \tau) |, \tau )
  )
  .
\end{align}
A straight-forward induction using~\eqref{eq:MHERGASsubdistri} and \eqref{eq:MHEInducStep} shows
\begin{align}
	e(t)
  \hspace{0.5cm}&\hspace{-0.5cm}
  \leq
  \kappa_{K}^{k} ( e(l) )
	\\
	& %
  \notag
	\plus
	\Plus _{n = 0} ^{k-1} [
  \kappa_{K}^{n} (
	\Plus _{\tau = 1} ^{K} [
  c( | w(t - (n K + \tau)) |, \tau )
  \\ & \hspace{2.5cm} \notag
	\plus
  d( | v(t - (n K + \tau)) |, \tau )
  ]
  )
  ]
  ,
\end{align}
which covers the time interval in which MHE is applied over horizon length ${K}$.
In order to include the starting interval as well, Theorem~\ref{thm:FIERGAS} and~\eqref{eq:MHERGASsubdistri} serve to obtain
\begin{align}
	e(t)
  \hspace{0.5cm}&\hspace{-0.5cm}
  \leq
  \kappa_{K}^{k} ( b( | x(0), \hat{x}(0) |, l ) )
	\\
	& \hspace{0.2cm}
  \notag
	\plus
	\Plus _{\tau = 1} ^{l}
  \kappa_{K}^{k} ( c( | w(l - \tau) |, \tau ) )
	\\
	& \hspace{0.2cm}
  \notag
	\plus
	\Plus _{\tau = 1} ^{l}
  \kappa_{K}^{k} ( d( | v(l - \tau) |, \tau ) )
	\\
	& \hspace{0.2cm}
  \notag
	\plus
	\Plus _{n = 0} ^{k-1} \Plus _{\tau = 1} ^{K}
  \kappa_{K}^{n} ( c( | w(t - (n K + \tau)) |, \tau ) )
	\\
	& \hspace{0.2cm}
  \notag
	\plus
	\Plus _{n = 0} ^{k-1} \Plus _{\tau = 1} ^{K}
  \kappa_{K}^{n} ( d( | v(t - (n K + \tau)) |, \tau ) )
  .
\end{align}
The desired estimate~\eqref{eq:MHERGASmax} results by defining
${\hat{b}_{K}}$, ${\hat{c}_{K}}$, and ${\hat{d}_{K}}$ according to~\eqref{eq:DefHatThetaK},
where the maximization serves to ensure that the defined functions are nonincreasing with respect to the second argument.
Finally, in the sum-based case, ${\hat{c}_{K}}$ and ${\hat{d}_{K}}$ are summable because of \eqref{eq:MHERGASsummable}.
\qed

If the horizon ${K}$ is chosen large enough, Theorem~\ref{thm:MHERGASmax} provides an RGAS result for MHE analogous to the FIE result in Theorem~\ref{thm:FIERGAS}.
Precisely, a) the MHE estimation error is bounded in terms of the initial estimation error and the actual process and output disturbance, b) convergence to zero of the disturbances implies the same for the estimation error, and c) explicit convergence rates are given in this case.
To the authors' best knowledge, an RGAS result for MHE has not been presented before without implying that the disturbance term ${\hat{c}_{K}}$ and ${\hat{d}_{K}}$ decrease exponentially with respect to their second argument.
Namely, previous robust stability results for MHE such as~\cite{Allan_Rawlings_MHE2019,Hu_arXiv_2017,Muller_ACC16,Muller_Aut_2017} effectively utilized a linear contraction ${\kappa_{K}}$ such that exponential convergence for convergent disturbances is implied.
However, there exist explicit RGES results for MHE in the literature that achieve a)-c) but require exponential i-IOSS and either a global Lipschitz condition for the system dynamics as in~\cite{knuefer2018} or an additional exponential incremental stabilizability condition with respect to the process disturbances (but no time-discounting), as in~\cite{Allan_Diss2020}.
As discussed in Remark~\ref{rem:MHERGES} below, exponential i-IOSS implies that there exists a \emph{linear} contraction ${\kappa_K}$ satisfying~\eqref{eq:MHERGASmaxcond}.
Since all linear contractions ${\kappa_K}$ satisfy~\eqref{eq:MHERGASsubdistri}-\eqref{eq:MHERGASsummable}, Theorem~\ref{thm:FIERGAS} covers the previous MHE results in~\cite{Allan_Rawlings_MHE2019,Hu_arXiv_2017,knuefer2018,Muller_ACC16,Muller_Aut_2017} as special cases and provides explicit statements about convergence for both the max-based and the sum-based formulation.

In general, FIE being RGAS might not necessarily provide a contraction map for MHE.
Precisely, there exist functions ${b}$ and ${\alpha}$ such that no horizon ${K}$ satisfies~\eqref{eq:MHERGASmaxcond}, e.g., ${b(r, t) = l(t)\sqrt{r}}$ and ${\alpha(r) = r}$ for any ${l \in \mathcal{L}}$.
In this case, extensions of Theorem~\ref{thm:MHERGASmax} are straight-forward that guarantee robust semi-global practical stability.\footnote{In contrast to global or local results, \emph{semi-global} results are valid on bounded, arbitrary large sets. \emph{Practical stability} (see, e.g., \cite[Definition~6.33]{GruenePannek_MPC_book_2017}) of the estimation error dynamics means that arbitrary small but a priori fixed neighborhoods of the origin are a) never left for sufficiently small initial estimation errors and b) approached arbitrarily close for increasing time under sufficiently small disturbances}
To this end, given that FIE is RGAS according to~\eqref{eq:FIERGAS}, we observe that~\eqref{eq:MHERGASmaxcond}-\eqref{eq:MHERGASsummable} can always be met for all ${r \in I}$ for arbitrary compact sets ${I \subset (0, \infty)}$ and for arbitrary linear contractions ${\kappa_K(r) = \eta_{K} r}$ with ${\eta_{K} \in (0, 1)}$, which results in exponentially decreasing functions ${\hat{b}_K(r,\cdot)}$, ${\hat{c}_K(r,\cdot)}$, and ${\hat{d}_K(r,\cdot)}$, see Remark~\ref{rem:MHERGES}.
These steps have been detailed in preprint~\cite{Hu_arXiv_2021} during the review process of the present work.
Note that although this restriction allows to obtain linear contractions ${\kappa_K}$, the resulting exponential decrease rates might be very slow such that nonlinear contractions might result in faster decrease rates and smaller gains.
In order to achieve robust semi-global asymptotic stability for some horizon ${K}$,~\eqref{eq:MHERGASmaxcond}-\eqref{eq:MHERGASsummable} need to be satisfied also close to the origin, i.e., for all ${r \in (0, R)}$ with arbitrary ${R \in (0, \infty)}$.
In the max-based case, this can always be guaranteed if there exists some ${K \in \mathbb{N}}$ such that ${b(\alpha^{-1}(\cdot), K)}$ is Lipschitz at the origin with Lipschitz constant ${L \leq 1}$.
Note that if and only if strict inequality, i.e., ${L < 1}$ holds for some ${K \in \mathbb{N}}$,~\eqref{eq:MHERGASmaxcond}-\eqref{eq:MHERGASsummable} are satisfied for a linear contraction ${\kappa_K}$ close to the origin. 
In Remark~\ref{rem:MHEcontractionImplicatons} below, sufficient conditions for the i-IOSS function ${\beta}$ are presented to guarantee that FIE according to Definition~\ref{defn:FIE} induces a contraction meeting the conditions of Theorem~\ref{thm:MHERGASmax}.
In addition to~\eqref{eq:MHERGASmaxcond}, the contraction ${\kappa_K}$ also needs to meet the relaxed distributivity inequality~\eqref{eq:MHERGASsubdistri} and \eqref{eq:MHERGASsummable}, which is guaranteed for all ${\mathcal{K}}$-functions in the max-based formulation.
For the sum-based formulation, \eqref{eq:MHERGASsubdistri}-\eqref{eq:MHERGASsummable} are not guaranteed for general ${\mathcal{K}}$-functions.
However, by slightly tightening condition~\eqref{eq:MHERGASmaxcond} and by sacrificing parts of the decrease rates, Theorem~\ref{thm:MHERGAS} below offers an alternative stability result in this case.

\begin{rem}
\label{rem:Application}
In order to apply the proposed MHE algorithm in practice, the i-IOSS condition should be checked first, for which a Lyapunov-based approach is described in~\cite{AllanRawlingsTeel_SIAM_2021,knuefer2020}.
In order to meet the compatibility condition in Assumption~\ref{ass:compatible}, the cost function can be designed according to Remark~\ref{rem:MainAssumption}.
The resulting FIE scheme satisfies~\eqref{eq:FIERGAS} by Theorem~\ref{thm:FIERGAS}.
In case the contraction conditions~\eqref{eq:MHERGASmaxcond}-\eqref{eq:MHERGASsummable} are not satisfied globally, one can either check whether other choices of the i-IOSS Lyapunov function allow to improve the i-IOSS estimates and hence reduce the composition ${b( \alpha^{-1}(r) , K )}$.
Alternatively, the relaxed (semi-global and/or practical stability) results discussed in the above paragraph might also be sufficient to apply MHE in practice.
\end{rem}

\begin{rem}
\label{rem:MHEcontractionImplicatons}
Assuming ${A = 1}$, i.e., optimal solutions in the MHE optimizations, and following Remark~\ref{rem:MainAssumption} and Proposition~\ref{prop:decrease}, the cost function~\eqref{eq:cost_K} can be designed compatible with the i-IOSS condition such that
\begin{align}
  \label{eq:MHEOptb}
  b(r, s)
  &=
  \begin{cases}
    \beta(2 r, s) & \text{max-based}
    \\
    2 \beta(\TriaConst{s} r, s) & \text{sum-based}
  \end{cases}
\end{align}
holds for all ${r \in [0, \infty)}$ and all ${s \in \mathbb{N}}$.
As discussed above, if there exists some ${K \in \mathbb{N}}$ such that ${b(\alpha^{-1}(\cdot), K)}$ satisfies a Lipschitz condition with Lipschitz constant ${L}$ sufficiently small, then~\eqref{eq:MHERGASmaxcond}-\eqref{eq:MHERGASsummable} can be satisfied also close to the origin, i.e., robust semi-global asymptotic stability can be guaranteed for MHE.
Due to the additional factors in~\eqref{eq:MHEOptb}, sufficient i-IOSS-conditions consequently impose tighter Lipschitz conditions for ${\beta}$.
To illustrate this, let us consider the max-based case and suppose ${\alpha^{-1}}$ has a Lipschitz constant of~${L_{\alpha^{-1}}}$ at the origin such that ${b(\cdot, K)}$ with Lipschitz constant ${L_{b} \leq 1/L_{\alpha^{-1}}}$ at the origin allows for robust semi-global asymptotic stability.
In this case, a sufficient condition in terms of ${\beta(\cdot, K)}$ requires a Lipschitz constant of~${L_{\beta} \leq 1/(2 L_{\alpha^{-1}})}$ at the origin.
To the authors' best knowledge, this is the least conservative condition in terms of the i-IOSS functions presented so far to ensure that MHE is robustly semi-globally asymptotically stable.\footnote{In~\cite[Assumption~1]{Allan_Rawlings_MHE2019}, a less conservative condition effectively allowing for any Lipschitz constant smaller than~${L_{\alpha^{-1}}^{-1}}$ is formulated, but in fact a condition equivalent to~\eqref{eq:OriginLipschitzBeta} is used in the proof of the according stability result, i.e.,~\cite[Theorem~1]{Allan_Rawlings_MHE2019}.}

In order to guarantee arbitrarily small linear contractions in the semi-global case, previous works, e.g.,~\cite{Hu_arXiv_2017,Muller_ACC16,Muller_Aut_2017}, imposed a stricter condition in terms of the i-IOSS function ${\beta}$, namely that there exist ${\sigma_{r} \in \mathcal{K}}$ Lipschitz continuous at the origin and ${\sigma_{s} \in \mathcal{L}}$ such that
\begin{align}
  \label{eq:OriginLipschitzBeta}
  \beta(r,s) \leq \sigma_{r}(r) \sigma_{s}(s)
\end{align}
holds for all ${r \in [0, \infty)}$ and all ${s \in \mathbb{N}}$, cf. the related discussion about the i-IOSS condition in~\cite{Allan_Rawlings_MHE2019}.
As above,~\eqref{eq:MHEOptb} shows that also ${b(\alpha^{-1}(\cdot), s)}$ satisfies an upper bound condition according to~\eqref{eq:OriginLipschitzBeta} if the cost function is chosen in a compatible way and if ${\alpha^{-1}}$ is Lipschitz at the origin.
For nonlinear observable systems and linear detectable systems, it is shown in~\cite{Ji_et_al_MHE} that the i-IOSS condition is fulfilled with ${\beta}$ satisfying~\eqref{eq:OriginLipschitzBeta}.
In~\cite{Allan_Rawlings_MHE2019}, this upper bound condition is discussed in detail and shown to imply a local exponential i-IOSS condition.
In fact, this is rather intuitive:
Firstly, a linear contraction guarantees MHE to be RGES, see Remark~\ref{rem:MHERGES}.
Secondly, existence of an RGES observer ensures the system to be exponentially i-IOSS, which is implied by~\cite[Proposition~2.6]{AllanRawlingsTeel_SIAM_2021}.

On the one hand,~\eqref{eq:OriginLipschitzBeta} implies the Lipschitz condition discussed at the beginning of this remark since the Lipschitz constant of ${\beta(\cdot, K)}$ at the origin can be rendered arbitrarily small for sufficiently large ${K}$.
On the other hand, any Lipschitz constant of ${\beta(\cdot, K)}$ at the origin strictly smaller than one implies that~\eqref{eq:OriginLipschitzBeta} holds semi-globally in ${r}$ and for all ${s \in \mathbb{N}}$, ${s \geq K}$, i.e., that the system admits a semi-global \emph{eventually\footnote{Naming according to~\cite{Allan_Rawlings_MHE2019}.}} exponential i-IOSS condition.

All in all, this remark depicts that a sufficiently small Lipschitz constant of ${\beta(\cdot, K)}$ at the origin implies a) that Proposition~\ref{prop:decrease} guarantees ${b}$ to satisfy the conditions of Theorem~\ref{thm:MHERGASmax} semi-globally and b) that the system admits a semi-global eventually exponential i-IOSS condition.
Note however that all conditions discussed in this remark are only sufficient stability conditions for MHE by Theorem~\ref{thm:MHERGASmax}, i.e., there might be less conservative approaches to guarantee that FIE induces a contraction than the one used in Proposition~\ref{prop:decrease}.
\end{rem}

\begin{rem}
\label{rem:MHERGES}
In case~\eqref{eq:MHERGASmaxcond} is satisfied for ${\kappa_{K}(r) = \eta_{K} r}$ with some ${\eta_{K} \in (0, 1)}$,~\eqref{eq:MHERGASsubdistri} directly applies with equality and also \eqref{eq:MHERGASsummable} is satisfied.
Furthermore, all functions ${\hat{b}_K}$, ${\hat{c}_K}$, and ${\hat{d}_K}$ decrease exponentially, i.e., can be bounded from above by ${C_{K} \lambda_{K}^{s} r}$ for some ${C_{K} \in [1, \infty)}$ and ${\lambda_{K} \in (0, 1)}$.
Hence, MHE is RGES if~\eqref{eq:MHERGASmaxcond} holds for linear contractions and if ${\alpha}$ equals the identity function.
Note that an exponential decrease of the estimation error is always guaranteed semi-globally under Lipschitz conditions for ${\beta}$ and ${\alpha^{-1}}$ as discussed above in Remark~\ref{rem:MHEcontractionImplicatons}.
\end{rem}

Theorem~\ref{thm:MHERGASmax} provides a stability result for all sufficiently large horizons ${K}$.
However, no direct statement is made how the resulting gains change for increasing horizons.
In fact, Theorem~\ref{thm:MHERGASmax} allows the contractions ${\kappa_{K}}$ to be independent of the horizon ${K}$.
Hence, the same relative decrease might only be achieved over increasing time intervals.
In case of ${\kappa_{K}}$ independent of ${K}$,
Definition~\eqref{eq:DefHatThetaK} implies
for instance ${\hat{b}_{K}(r, K) = \kappa_K(b(r, 0))}$, i.e., that in general the convergence rates can neither be rendered arbitrarily close to the FIE rates nor bounded from above uniformly in ${K}$ by ${\mathcal{KL}}$-functions.\footnote{Cf. the related discussion in Remark~11 in \cite{Muller_Aut_2017}.}
Both can be guaranteed in the max-based case according to the following statement about improving gains.

\begin{thm}
\label{thm:ImprovingBounds}
Consider the max-based formulation and suppose FIE satisfies~\eqref{eq:FIERGAS}.
If there exists ${K_0 \in \mathbb{N}}$ such that
\begin{align}
  \label{eq:MHERGASmaxcond2}
  b( \alpha^{-1}(r) , K_0 ) <  r
\end{align}
holds for all ${r \in (0, \infty)}$, then MHE is RGAS for all horizons ${K \geq K_0}$ and the corresponding gains approach the FIE gains for increasing horizons.
Precisely, for ${\theta \in \{b, c, d\}}$, there exist sequences of functions ${\{\bar{\theta}_{K}\}_{K \in \mathbb{N}}}$, ${\bar{\theta}_{K} \in \mathcal{KL}}$, such that a) \eqref{eq:MHERGASmax} is satisfied for ${K \geq K_0}$, b) ${\bar{\theta}_{K} \rightarrow \theta}$ pointwise for ${K \rightarrow \infty}$, and c) ${\bar{\theta}_{K}(r, t) \geq \bar{\theta}_{K+1}(r, t)}$ for all ${K, t \in \mathbb{N}}$ and ${r \in [0, \infty)}$.
\end{thm}

\textbf{Proof of Theorem~\ref{thm:ImprovingBounds}:}
For all ${K \in \mathbb{N}}$, define ${\hat{b}_{K}}$, ${\hat{c}_{K}}$, and ${\hat{d}_{K}}$ according
to~\eqref{eq:DefHatThetaK} with ${\kappa_{K}(r) := b( \alpha^{-1}(r) , K )}$.
Since all arguments apply for ${b}$, ${c}$, and ${d}$ alike, we present them once using the placeholders ${\bcd \in \{b, c, d\}}$ and ${\hat{\bcd}_{k} \in \{\hat{b}_{k}, \hat{c}_{k}, \hat{d}_{k}\}}$.
Based on~\eqref{eq:DefHatThetaK}, we define the sequence ${\{\bar{\bcd}_{K}\}_{K \in \mathbb{N}}}$ by
\begin{align}
  \label{eq:DefBarbK}
  \bar{\bcd}_{K}(r, t) &:=
  \begin{cases}
  \sup_{k \in \mathbb{N}, k \geq K} \hat{\bcd}_{k}( r, t )
  & \text{for } K \geq K_0
  \\\sup_{k \in \mathbb{N}, k \geq K_0} \hat{\bcd}_{k}( r, t )
  & \text{for } K < K_0,
  \end{cases}
\end{align}
where ${\hat{\bcd}_{K}( r, t ) \leq \hat{\bcd}_{K}( r, 0 ) = \bcd(r, 0)}$
for all ${K, t \in \mathbb{N}}$, ${r \in [0, \infty)}$ implies well-definedness, i.e., that the suprema in~\eqref{eq:DefBarbK} exist in $[0, \infty)$.
By definition, for all ${K \in \mathbb{N}}$ with ${K \geq K_0}$, the conditions of Theorem~\ref{thm:MHERGASmax} apply, see Footnote~\ref{fnt:RGASCondMaxBased}, and \eqref{eq:MHERGASmax} is satisfied with the triple ${\bar{b}_{K}}$, ${\bar{c}_{K}}$, and ${\bar{d}_{K}}$, which gives a).

For the remainder of this proof, fix ${t \in \mathbb{N}}$ and ${r \in [0, \infty)}$ arbitrary.
Since by definition ${\bar{\bcd}_{K} = \max\{ \hat{\bcd}_{K}, \bar{\bcd}_{K+1}\}}$ and consequently ${\bar{\bcd}_{K}(r, t) \geq \bar{\bcd}_{K+1}(r, t)}$ hold for all ${K \in \mathbb{N}}$, ${K \geq K_0}$, it is straight-forward to show c).

Since ${\kappa_K \rightarrow 0}$ pointwise for ${K \rightarrow \infty}$, there exists ${\bar{k} \in \mathbb{N}}$ such that ${\kappa_{K}(\bcd(r, 0)) < \bcd(r, t)}$ applies for all ${K \in \mathbb{N}}$, ${K > \bar{k}}$.
Consequently, for all ${K \geq \bar{K} := \max \{\bar{k}, t\}}$, we have ${\hat{\bcd}_{K}(r, t) = \bcd(r, t)}$ because ${\lfloor t/K \rfloor = 0}$ and ${t \mod K = t}$.
Hence, also ${\bar{\bcd}_{K}(r, t) = \bcd(r, t)}$ holds for all ${K \in \mathbb{N}}$, ${K \geq \bar{K}}$.
Since ${t}$ and ${r}$ are arbitrary, this implies ${\bar{\bcd}_{K} \rightarrow \bcd}$ pointwise for ${K \rightarrow \infty}$, i.e., b).
Since ${\hat{\bcd}_{K} \in \mathcal{KL}}$ holds for all ${K \in \mathbb{N}}$, also ${\bar{\bcd}_{K}(0, t) = 0}$, ${\bar{\bcd}_{K}(r, t) \geq 0}$, and ${\bar{\bcd}_{K}(r, \cdot)}$ nonincreasing follow.
It remains to show that, for all ${K \in \mathbb{N}}$, ${\bar{\bcd}_{K}}$ is continuous, approaches zero for increasing second arguments, and is unbounded for increasing first arguments.
For sufficiently large ${K}$, these three properties follow as a neighborhood of the pointwise limit ${\bcd}$ is reached.
For all smaller ${K}$, they follow by induction, which concludes the proof.
\qed

\begin{rem}
\label{rem:decreaseratesSUM}
In the max-based case, Theorem~\ref{thm:ImprovingBounds} provides a) decrease rates for the RGAS estimate~\eqref{eq:MHERGASmax} of MHE for each horizon ${K}$, which b) approach the corresponding FIE decrease rates for increasing horizons, and which c) give an upper bound for all decrease rates obtained for larger horizons.
Since maximization is always distributive with respect to ${\mathcal{K}}$-functions according to \eqref{eq:MHERGASsubdistri} and since \eqref{eq:MHERGASsummable} trivially applies, only~\eqref{eq:MHERGASmaxcond2} needs to be assumed.
Also in the sum-based case, we can guarantee a) and b) (with summable decrease rates) by defining ${\bar{\bcd}_{K} := \hat{\bcd}_{K}}$ if there exists a sequence of contractions ${\kappa_{K}}$ satisfying~\eqref{eq:MHERGASmaxcond}-\eqref{eq:MHERGASsummable} such that ${\kappa_K \rightarrow 0}$ pointwise for ${K \rightarrow \infty}$.
In order to achieve c) as well while maintaining summability of ${\bar{\bcd}_{K}}$, additional technical conditions about the convergence rate of ${\kappa_K \rightarrow 0}$ for ${K \rightarrow \infty}$ are needed.
\end{rem}

As discussed above of Remark~\ref{rem:Application}, conditions~\eqref{eq:MHERGASsubdistri}-\eqref{eq:MHERGASsummable} of Theorem~\ref{thm:MHERGASmax} might in general not hold in the sum-based case.
Nevertheless, the following slightly weaker stability result can be made without relying on~\eqref{eq:MHERGASsubdistri}-\eqref{eq:MHERGASsummable}.

\begin{thm}[RGAS of MHE, ${\plus = +}$]
\label{thm:MHERGAS}
Consider the sum-based formulation and suppose FIE satisfies~\eqref{eq:FIERGAS}.
If for the horizon length ${K}$ there exist ${\kappa_{K}, \rho_{K} \in \mathcal{K}_{\infty}}$ such that
\begin{align}
  \label{eq:MHERGAScond}
  b( \alpha^{-1}(r) , K ) + \rho_{K}(r) \leq \kappa_{K}(r) < r
\end{align}
holds for all ${r \in (0, \infty)}$, then MHE is RGAS in the following sense.
The state estimate resulting from MHE satisfies
\begin{align}
	\label{eq:MHERGAS}
	\alpha(|x(t), \hat{x}(t)|)
  \leq
  & \max \{ \hat{b}_{K}( | x(0), \hat{x}(0) |, t ),
	\\
	&
  \notag
  \max_{1 \leq \tau \leq t} \hat{c}_{K}( | w(t - \tau) |, \tau ),
  \\
	&
  \notag
  \max_{1 \leq \tau \leq t} \hat{d}_{K}( | v(t - \tau) |, \tau ) \}
\end{align}
for all ${t \in \mathbb{N}}$, all ${\hat{x}_{0|0} \in \Xest}$, ${x_0 \in \Xcon}$, ${w \in \Wcon^{\infty}}$, and ${v \in \Vcon^{\infty}}$
with ${\hat{b}_{K}, \hat{c}_{K}, \hat{d}_{K} \in \mathcal{KL}}$ according to
\begin{align}
  \label{eq:DefHatbKsum}
  \hat{b}_{K}(r, t) &:=
  \max \{ \kappa_{K}^{\lfloor t / K \rfloor} ( 2 b( r, t \bmod K )),
  \\
  \notag
  & \hspace{1.35cm}
  \kappa_{K}^{\lfloor t / K \rfloor + 1} ( 2 b( r, 0 )) \}
  \\
  \label{eq:DefHatcKsum}
  \hat{c}_{K}(r, t) &:=
  \kappa_{K}^{\lfloor t / K \rfloor} ( \zeta_{K} ( 2 \sum _{\tau = 1} ^{K} c( r, \tau ) ) )
  \\
  \label{eq:DefHatdKsum}
  \hat{d}_{K}(r, t) &:=
  \kappa_{K}^{\lfloor t / K \rfloor} ( \zeta_{K} ( 2 \sum _{\tau = 1} ^{K} d( r, \tau ) ) )
\end{align}
with ${\zeta_{K}(r) := r + \kappa_{K} \circ \rho_{K}^{-1} (r) \in \mathcal{K}_{\infty}}$.
\end{thm}

\textbf{Proof of Theorem~\ref{thm:MHERGAS}:}
Let ${e(t) := \alpha(|x(t), \hat{x}(t)|)}$ and ${t = k K + l}$ with ${k, l \in \mathbb{N}}$ and minimal ${l}$, i.e., ${l = t \bmod K}$ and ${k = \lfloor t / K \rfloor}$.
Since the MHE optimization problem is derived from FIE, we can directly apply \eqref{eq:FIERGAS} of Theorem~\ref{thm:FIERGAS} and~\eqref{eq:MHERGAScond} to obtain
\begin{align}
	e(t)
  &\leq
  \label{eq:PREiterstepK}
  \kappa_{K} (e(t-K))
  - \rho_{K} (e(t-K))
	\\
	& \hspace{0.5cm}
  \notag
  + \sum _{\tau = 1} ^{K} ( c( | w(t - \tau) |, \tau )
	+ d( | v(t - \tau) |, \tau ) )
  .
\end{align}
Using the definition of ${\zeta_{K}}$ below~\eqref{eq:DefHatdKsum} and a distinction of cases as in the proof of~\cite[Proposition~5]{Allan_Rawlings_ACC2019} (or similarly of~\cite[Theorem~9]{Muller_Aut_2017}), we obtain
\begin{align}
  \label{eq:iterstepK}
	e(t)
  &\leq
  \max \{
  \kappa_{K} (e(t-K)),
  \\
	& \hspace{0.5cm}
  \notag
  \zeta_{K}(\sum _{\tau = 1} ^{K} [ c( | w(t - \tau) |, \tau )
	+ d( | v(t - \tau) |, \tau ) ] )
  \},
\end{align}
i.e., a max-based nonlinear contraction instead of the sum-based one in~\eqref{eq:PREiterstepK}.
Hence, as in the proof of Theorem~\ref{thm:MHERGASmax}, a straight-forward induction and Theorem~\ref{thm:FIERGAS} for the starting interval yield
\begin{align}
  \label{eq:MHERGASsumIntermediate}
	e(t)
  \leq
  & \max \{
  \kappa_{K}^{k} ( b( | x(0), \hat{x}(0) |, l )
	\\
	& \hspace{0.5cm}
  \notag
  + \sum _{\tau = 1} ^{l} [ c( | w(l - \tau) |, \tau )
	+ d( | v(l - \tau) |, \tau ) ] ),
	\\
	& \hspace{0.3cm}
  \notag
  \max_{0 \leq n < k} \kappa_{K}^{n} ( \zeta_{K} (
  \sum _{\tau = 1} ^{K} [ c( | w(t - (n K + \tau)) |, \tau )
	\\
	& \hspace{0.5cm}
  \notag
	+ d( | v(t - (n K + \tau)) |, \tau ) ]
  ) )
  \}.
\end{align}
In order to obtain distinct gains for the disturbances of each time instant, we would need to pull the ${\kappa_{K}^{\cdot}}$-terms into the sums.
Since this does in general not allow to master the resulting upper bounds, we again transfer to a max-based formulation by observing that
\begin{align}
  \label{eq:ConservativeSumBound}
  \sum _{\tau = 1} ^{T} \kappa( r_{\tau}, \tau )
  \leq
  \max_{1 \leq \tau \leq T} \sum _{s = 1} ^{T} \kappa( r_{\tau}, s )
\end{align}
applies for all sequences ${\{r_{\tau}\}_{\tau \in \mathbb{N}}}$,  ${r_{\tau} \in [0, \infty)}$, all ${T \in \mathbb{N}}$, and all summable ${\kappa \in \mathcal{KL}}$.
Using this approach and noting ${\zeta_{K}(r) > 2r}$ finally yields
\begin{align}
	e(t)
  \leq
  \label{eq:thm:MHERGASproofresult}
  & \max \Big\{ \kappa_{K}^{k} ( 2 b( | x(0), \hat{x}(0) |, l )),
	\\
	&
  \notag
  \max_{0 \leq n < k} \max _{1 \leq \tau \leq K} \kappa_{K}^{n} ( \zeta_{K} ( 2 c_{K}( | w(t - (n K + \tau)) | ) ) ),
  \\
	&
  \notag
  \max_{0 \leq n < k} \max _{1 \leq \tau \leq K} \kappa_{K}^{n} ( \zeta_{K} ( 2 d_{K}( | v(t - (n K + \tau)) | ) ) ),
	\\
	&
  \notag
  \max _{1 \leq \tau \leq l} \kappa_{K}^{k} ( \zeta_{K} ( 2 c_{K}( | w(t - (k K + \tau)) | ) ) ),
  \\
	&
  \notag
  \max _{1 \leq \tau \leq l} \kappa_{K}^{k} ( \zeta_{K} ( 2 d_{K}( | v(t - (k K + \tau)) | ) ) ) \Big\}
\end{align}
with
\begin{align}
  \label{eq:abbrev:csumK}
  c_{K}(r)
  &
  := \sum_{\tau = 1} ^{K} c(r, \tau)
  \quad
  \text{and}
  \quad
  d_{K}(r)
  := \sum_{\tau = 1} ^{K} d(r, \tau),
\end{align}
which is the stated stability result.
\qed

The key idea in the above proof is to transfer the sum-based contraction to a max-based contraction in~\eqref{eq:iterstepK} using ${\zeta_{K}}$.
This allows to iterate without using~\eqref{eq:MHERGASsubdistri} as done in Theorem~\ref{thm:MHERGASmax}.
The resulting estimate in~\eqref{eq:MHERGASsumIntermediate} would already give a certain kind of stability estimate that differs from the structure of~\eqref{eq:MHERGAS}, as the disturbance terms are averaged over the horizon length.
This averaging is eliminated by the conservative relaxation according to~\eqref{eq:ConservativeSumBound}, which causes the structural difference of the disturbance gains defined in~\eqref{eq:DefHatcKsum}-\eqref{eq:DefHatdKsum} compared to the ones in Theorem~\ref{thm:MHERGASmax}.
Note that although the inner discounting via ${c}$ and ${d}$ is sacrificed compared to~\eqref{eq:DefHatThetaK}, \eqref{eq:DefHatcKsum}-\eqref{eq:DefHatdKsum} still define ${\mathcal{KL}}$-functions due to the iterated function ${\kappa_{K}^{\lfloor t/K \rfloor}}$.
Moreover, the sums in these definitions do indeed increase for larger horizons but remain bounded if ${c}$ and ${d}$ are summable as assumed for the sum-based case.
Finally, we observe that although a sum-based MHE formulation is considered in Theorem~\ref{thm:MHERGAS}, a max-based stability result is obtained due to the max-based iteration in the proof.
This perfectly agrees with the previous RGES results for MHE, where non time-discounted max-based results are presented, see~\cite{Allan_Rawlings_MHE2019,Hu_arXiv_2017,Hu_et_al_CDC15,Muller_ACC16,Muller_Aut_2017}.
As discussed in~\cite[Chapter~2.2]{Allan_Diss2020} and~\cite[Remark~4]{knuefer2020}, for exponentially time-discounted estimates the max-based structure in~\eqref{eq:MHERGAS} is equivalent to the sum-based one used for instance in RGES results for FIE and MHE in~\cite{knuefer2020} and in~\cite[Theorems~5.28 and 5.30]{Allan_Diss2020}.
In case of Theorem~\ref{thm:MHERGAS}, the max-based result can simply be relaxed to a sum-based result if \eqref{eq:MHERGASsummable} is satisfied additionally, i.e., if the decrease resulting form the contraction ${\kappa_{K}}$ ensures ${\hat{c}_{K}}$ and ${\hat{d}_{K}}$ to be summable.

\begin{rem}
\label{rem:MHEsumInnerDiscounting}
If the function ${c}$ is of a recursive form, e.g., ${c(r, t) = \kappa_{c}^{t}(\sigma_{c}(r))}$ with ${\kappa_{c}, \sigma_{c} \in \mathcal{K}_{\infty}}$\footnote{Note that this recursive form results from the Lyapunov-like condition for i-IOSS discussed in~\cite{knuefer2020} and a cost function according to Remark~\ref{rem:MainAssumption}.}, the conservative step~\eqref{eq:ConservativeSumBound} can be improved to
\begin{align}
  \label{eq:SumBoundDiscountingPreserved}
  \sum _{\tau = 1} ^{T} c( r_{\tau}, \tau )
  &\leq
  \max_{1 \leq \tau \leq T} \{ \sum _{s = 1} ^{T} \kappa_{c}^{\lfloor s/2 \rfloor}( c(r_{\tau}, \lfloor \tau/2 \rfloor) ) \}
  \\
  &=:
  \max_{1 \leq \tau \leq T} \{ \tilde{c}_{T} (r_{\tau}, \tau) \}
\end{align}
for all sequences ${\{r_{\tau}\}_{\tau \in \mathbb{N}}}$,  ${r_{\tau} \in [0, \infty)}$ and all ${T \in \mathbb{N}}$ with ${\tilde{c}_{T} \in \mathcal{KL}}$. %
Hence, the terms ${c_{K}(|w(t-\tau)|)}$ in~\eqref{eq:thm:MHERGASproofresult} can be replaced by terms of the form ${\tilde{c}_{K} (|w(t-\tau)|, \tau)}$.
Since the same arguments apply for ${d}$, the inner discounting over the horizon ${K}$ can be preserved for the disturbances ${w}$ and ${v}$ in Theorem~\ref{thm:MHERGAS}.
Note that the above steps offer a general approach to transfer discounted sums into discounted max-based terms for recursively defined ${\mathcal{KL}}$-functions, cf. \cite[Remark~4]{knuefer2020}.
\end{rem}

\begin{rem}
\label{rem:MHENonDiscounted}
Remarks~\ref{rem:NonDiscounted} and~\ref{rem:FIENonDiscounted} above discuss a non time-discounted setup for Proposition~\ref{prop:decrease} and Theorem~\ref{thm:FIERGAS}.
For Theorems~\ref{thm:MHERGASmax} and~\ref{thm:MHERGAS}, the results apply unchanged in this case.
In Theorem~\ref{thm:MHERGASmax}, the disturbance gains ${\hat{c}_{K}}$ and ${\hat{d}_{K}}$ remain bounded for increasing horizons ${K}$ for such a setup.
In contrast to this, the according gains increase for larger ${K}$ in Theorem~\ref{thm:MHERGAS} due to the sums in~\eqref{eq:DefHatcKsum}-\eqref{eq:DefHatdKsum}.
Similar observations are made in~\cite[Remarks~16 and~18]{Muller_Aut_2017}.
\end{rem}

\section{Conclusions}
\label{sec:conclusions}
The contributions of this work concern optimization-based state estimation in terms of FIE and MHE but provide a fundamental result about nonlinear detectability as well:
The RGAS result for FIE in Theorem~\ref{thm:FIERGAS} effectively only\footnote{Note that the time-discounted cost function needs to be compatible with the i-IOSS estimate, but this can always be guaranteed by a proper design, see Remark~\ref{rem:MainAssumption}.} relies on i-IOSS and hence shows that this detectability condition is indeed sufficient for the existence of RGAS observers.
Moreover, convergence of the estimation error is guaranteed in case of convergent disturbances without any additional assumptions.
In case the FIE stability result induces a strict (not necessarily linear) global contraction of the estimation error over a sufficiently large horizon, the induced MHE scheme is RGAS as well according to Theorem~\ref{thm:MHERGASmax} and Theorem~\ref{thm:MHERGAS}.
For both FIE and MHE, explicit convergence rates are presented in these results.
Furthermore, Theorem~\ref{thm:ImprovingBounds} shows under which conditions the MHE results approach the FIE results for increasing horizons.
Finally, we discuss in detail in which aspects these results use less conservative conditions compared to previous MHE and FIE results.

Robust stability and convergence of FIE is fully established by Theorem~\ref{thm:FIERGAS}, cf. the discussion in~\cite[Chapter~6]{Allan_Diss2020}.
Although similar stability results for MHE are presented in this work, the crucial step still requires FIE to induce a global contraction for the estimation error.
As discussed in Remark~\ref{rem:MHEcontractionImplicatons}, our current arguments to guarantee such a global contraction require an \emph{eventually} exponential i-IOSS condition to hold.
However, all above results can directly be generalized towards robust semi-global practical stability which circumvents this implication.
The step from FIE to MHE, i.e., how to relax the implication of an \emph{eventually} exponential i-IOSS condition, is subject to ongoing research.


\begin{thebibliography}{10}

\bibitem{Alessandri_et_al_Aut08}
A.~Alessandri, M.~Baglietto, and G.~Battistelli.
\newblock Moving-horizon state estimation for nonlinear discrete-time systems:
  New stability results and approximation schemes.
\newblock {\em Automatica}, 44(7):1753 --1765, 2008.

\bibitem{Alessandri_et_al_CDC10}
A.~Alessandri, M.~Baglietto, G.~Battistelli, and V.~Zavala.
\newblock Advances in moving horizon estimation for nonlinear systems.
\newblock In {\em 49th IEEE Conference on Decision and Control (CDC)}, pages
  5681--5688, 2010.

\bibitem{Allan_Diss2020}
D.~A. Allan.
\newblock {\em A Lyapunov-Like Function for Analysis of Model Predictive
  Control and Moving Horizon Estimation}.
\newblock {P}h{D} thesis, University of Wisconsin-Madison, 2020.

\bibitem{Allan_Rawlings_ACC2019}
D.~A. Allan and J.~B. Rawlings.
\newblock A {L}yapunov-like function for full information estimation.
\newblock In {\em American Control Conference (ACC)}, pages 4497--4502, 2019.

\bibitem{Allan_Rawlings_MHE2019}
D.~A. Allan and J.~B. Rawlings.
\newblock Moving horizon estimation.
\newblock In Sa{\v{s}}a~V. Rakovi{\'{c}} and William~S. Levine, editors, {\em
  Handbook of Model Predictive Control}, pages 99--124. Springer, Cham, 2019.

\bibitem{AllanRawlingsTeel_SIAM_2021}
D.~A. Allan, J.~B. Rawlings, and A.~R. Teel.
\newblock Nonlinear detectability and incremental input/output-to-state
  stability.
\newblock {\em SIAM Journal on Control and Optimization}, 59(4):3017--3039,
  2021.

\bibitem{Gharbi_Ebenbauer_2019}
M.~Gharbi and C.~Ebenbauer.
\newblock A proximity moving horizon estimator based on {B}regman distances and
  relaxed barrier functions.
\newblock In {\em Proc.\ 18th European Control Conference (ECC)}, pages
  1790--1795, Napoli, Italy, 2019.

\bibitem{Gharbi_Ebenbauer_2020}
M.~Gharbi and C.~Ebenbauer.
\newblock A proximity moving horizon estimator for a class of nonlinear
  systems.
\newblock {\em International Journal of Adaptive Control and Signal
  Processing}, 34(6):721--742, 2020.

\bibitem{GruenePannek_MPC_book_2017}
L.~Gr{\"u}ne and J.~Pannek.
\newblock {\em Nonlinear Model Predictive Control}.
\newblock Springer International Publishing, Cham, 2017.

\bibitem{Hu_arXiv_2017}
W.~Hu.
\newblock Robust stability of optimization-based state estimation.
\newblock {\em arXiv preprint arXiv:1702.01903}, 2017.

\bibitem{Hu_arXiv_2021}
W.~Hu.
\newblock Generic stability implication from full information estimation to
  moving-horizon estimation.
\newblock {\em arXiv preprint arXiv:2105.10125}, 2021.

\bibitem{Hu_et_al_CDC15}
W.~Hu, L.~Xie, and K.~You.
\newblock Optimization-based state estimation under bounded disturbances.
\newblock {\em Proceedings of the 54th IEEE Conference on Decision and
  Control}, pages 6597--6602, 2015.

\bibitem{Ji_et_al_MHE}
L.~Ji, J.~B. Rawlings, W.~Hu, A.~Wynn, and M.~Diehl.
\newblock Robust stability of moving horizon estimation under bounded
  disturbances.
\newblock {\em IEEE Transactions on Automatic Control}, 61(11):3509--3514,
  2016.

\bibitem{knuefer2018}
S.~Kn\"ufer and M.~A. M\"uller.
\newblock Robust global exponential stability for moving horizon estimation.
\newblock In {\em 57th IEEE Conference on Decision and Control (CDC)}, pages
  3477--3482. IEEE, 2018.

\bibitem{knuefer2020}
S.~Kn{\"u}fer and M.~A M{\"u}ller.
\newblock Time-discounted incremental input/output-to-state stability.
\newblock In {\em 59th IEEE Conference on Decision and Control (CDC)}, pages
  5394--5400. IEEE, 2020.

\bibitem{Michalska_Mayne_1992}
H.~Michalska and D.~Q. Mayne.
\newblock Moving horizon observers.
\newblock {\em IFAC Proceedings Volumes}, 25(13):185--190, 1992.

\bibitem{Muller_ACC16}
M.~A. M\"uller.
\newblock Nonlinear moving horizon estimation for systems with bounded
  disturbances.
\newblock In {\em Proceedings of the American Control Conference}, pages
  883--888, 2016.

\bibitem{Muller_Aut_2017}
M.~A. M{\"u}ller.
\newblock Nonlinear moving horizon estimation in the presence of bounded
  disturbances.
\newblock {\em Automatica}, 79:306 -- 314, 2017.

\bibitem{Muske_et_al_ACC_1993}
K.~R. Muske, J.~B. Rawlings, and J.~H. Lee.
\newblock Receding horizon recursive state estimation.
\newblock In {\em American Control Conference, 1993}, pages 900--904. IEEE,
  1993.

\bibitem{Rao_et_al_TAC03}
C.~V. Rao, J.~B. Rawlings, and D.~Q. Mayne.
\newblock Constrained state estimation for nonlinear discrete-time systems:
  stability and moving horizon approximations.
\newblock {\em IEEE Transactions on Automatic Control}, 48(2):246--258, 2003.

\bibitem{Rawlings_Ji_JPC12}
J.~B. Rawlings and L.~Ji.
\newblock Optimization-based state estimation: Current status and some new
  results.
\newblock {\em Journal of Process Control}, 22:1439--1444, 2012.

\bibitem{Rawlings_Mayne_09}
J.~B. Rawlings and D.~Q. Mayne.
\newblock {\em {Model Predictive Control: Theory and Design}}.
\newblock Nob Hill Publishing, Madison, WI, 2009.

\bibitem{Rawlings_Mayne_Diehl_MPC17}
J.~B. Rawlings, D.~Q. Mayne, and M.~Diehl.
\newblock {\em Model predictive control: theory, computation, and design},
  volume~2.
\newblock Nob Hill Publishing Madison, WI, 2017.

\bibitem{Sontag_1998}
E.~D. Sontag.
\newblock Comments on integral variants of {ISS}.
\newblock {\em Systems \& Control Letters}, 34(1):93--100, 1998.

\bibitem{Sontag_OSS}
E.~D. Sontag and Y.~Wang.
\newblock Output-to-state stability and detectability of nonlinear systems.
\newblock {\em Systems \& Control Letters}, 29(5):279--290, 1997.

\bibitem{Wynn_et_al_TAC14}
A.~Wynn, M.~Vukov, and M.~Diehl.
\newblock Convergence guarantees for moving horizon estimation based on the
  real-time iteration scheme.
\newblock {\em IEEE Transactions on Automatic Control}, 59(8):2215--2221, 2014.

\end{thebibliography}
\end{document}